\documentclass[conference]{IEEEtran}
\IEEEoverridecommandlockouts
\usepackage{cite}
\usepackage{amsmath,amssymb,amsfonts}
\usepackage{algorithmic}
\usepackage{graphicx}
\usepackage{textcomp}
\usepackage{xcolor}
\def\BibTeX{{\rm B\kern-.05em{\sc i\kern-.025em b}\kern-.08em
    T\kern-.1667em\lower.7ex\hbox{E}\kern-.125emX}}
\begin{document}

\title{Power-Integrity Modeling of VR Faults in High-Performance Applications

\thanks{This work was supported by the Center for Heterogeneous Integration of Micro Electronic Systems (CHIMES), one of seven centers in Joint University Microelectronics Program (JUMP) 2.0, a Semiconductor Research Corporation (SRC) program sponsored by the Defense Advance Research Project Agency (DARPA).}
}

\author{\IEEEauthorblockN{Sriharini Krishnakumar}
\IEEEauthorblockA{\textit{Electrical and Computer Engineering} \\
\textit{University of Illinois Chicago}\\
Chicago, IL, USA \\
skrish47@uic.edu}
\and
\IEEEauthorblockN{Inna Partin-Vaisband}
\IEEEauthorblockA{\textit{Electrical and Computer Engineering} \\
\textit{University of Illinois Chicago}\\
Chicago, IL, USA \\
vaisband@uic.edu}
}

\maketitle

\begin{abstract}
Distributed vertical power delivery has emerged as a promising approach to meet aggressive current-density, efficiency, and transient-response requirements in high-performance computing systems. Tight integration of voltage regulators within stacked substrates, however, increases the vulnerability of the power delivery system to short-circuit and open-circuit faults arising from elevated thermal and mechanical stresses. Such faults can propagate through the shared power delivery network, leading to rapid degradation of system-wide efficiency at worst-case rates of up to 0.5\% per microsecond. Advanced fault-tolerant power management strategies are therefore required to ensure efficient power delivery. A real-time fault-detection and isolation methodology are proposed in this paper for vertical power delivery systems. The methodology is developed based on an analytical inductor-current models that rely solely on signals available within the converter control circuitry, thereby eliminating additional sensing overhead. The proposed framework is designed and simulated in SPICE environment, demonstrating sub-microsecond fault detection and effective dual-fuse isolation, maintaining uninterrupted power delivery with a system-wide efficiency degradation of less than 2\%.

\end{abstract}


\section{Introduction}
\begin{figure}[t!]
\vspace{-10pt}
\centerline{\includegraphics[width=0.45\textwidth]{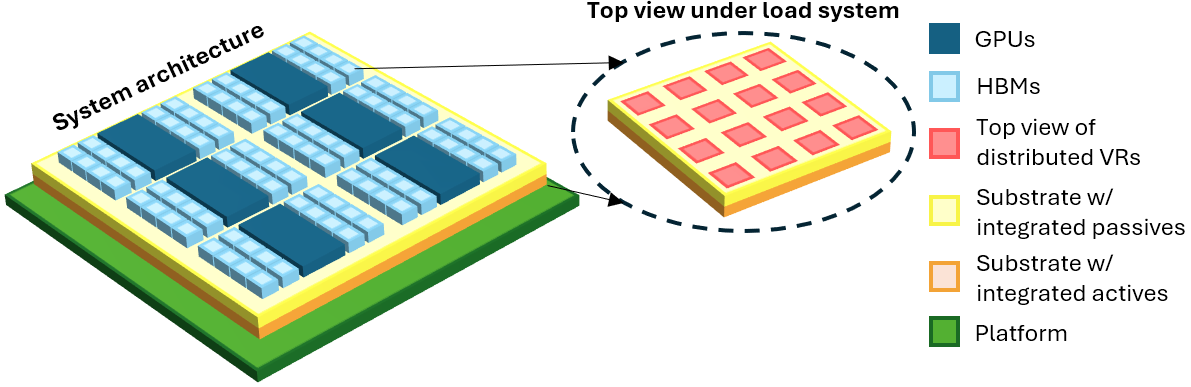}}
\caption{Schematic of a DVPD system.}
\label{fig:DVPD}
\vspace{-15pt}
\end{figure}

The rapid advancement of artificial intelligence (AI) workloads has significantly increased the demand for high-performance processors and accelerators, which require high current densities to support large-scale data processing and complex model training. Efficient power delivery in these systems is enabled by advanced power distribution architectures that employ laterally distributed voltage regulators (VRs) with vertically stacked components, commonly referred to as distributed vertical power delivery (DVPD) \cite{krishnakumar2023vertical,krishnakumar2024system}. In a DVPD system, as illustrated in Fig. \ref{fig:DVPD}, multiple VRs are connected in parallel to a common low-voltage power plane that supplies a large number of spatially distributed heterogeneous loads, such as graphics processing units (GPUs) and high-bandwidth memories (HBMs), mounted on a shared substrate. Each VR delivers a fraction of the total load current, enabling current sharing across multiple parallel conduction paths. This architecture reduces the electrical distance between the VRs and points of load (POLs), lowers parasitic resistance and loop inductance, and enables fast transient response while supporting high power-density operation.

However, the increasing integration density and architectural complexity of DVPD systems elevates thermal, electrical, and mechanical stresses within the power delivery system, thereby increasing the likelihood of component-level failures. The strong electrical coupling between the converter power stages, the power delivery network (PDN), and the load system allows non-idealities or faults within an individual VR to propagate through the shared PDN and disrupt voltage regulation across the entire system. Consequently, system-level reliability has emerged as a critical concern in DVPD architectures and remains insufficiently investigated. Maintaining uninterrupted power delivery to critical loads under fault conditions therefore requires effective fault-tolerant power-management strategies. In general, a fault-tolerant power system must provide three essential capabilities: VR redundancy, fault detection, and fault isolation. Although DVPD architectures inherently support redundancy, effective fault detection and isolation mechanisms are required to prevent fault propagation through the PDN and preserve system-level performance.
\begin{figure*}[t!]
\centerline{\includegraphics[width=0.8\textwidth]{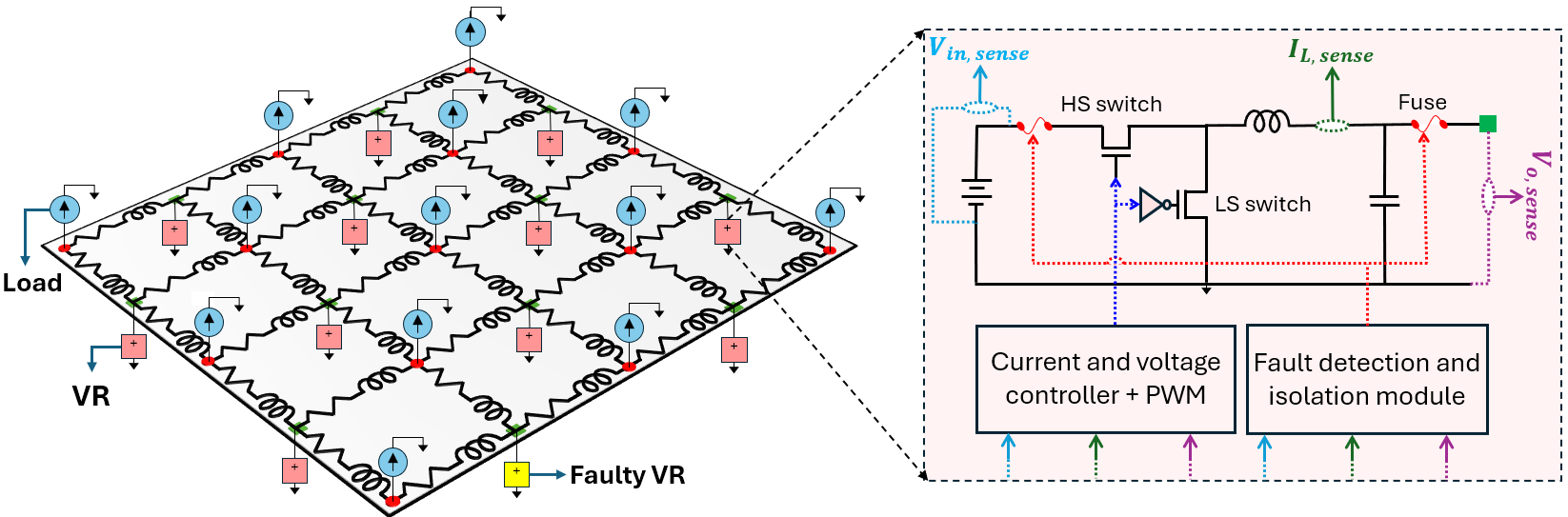}}
\vspace{-5pt}
\caption{Illustration of DVPD system comprising multiple parallel VRs connected to a shared PDN. The inset shows a representative buck converter with integrated current and voltage sensing, control, and pulse-width modulator (PWM), along with the proposed fault-detection and isolation module. Input voltage, inductor current, and sensed output voltage are used by the fault-detection module to enable fast fault identification and isolation.}
\label{fig:PDN}
\vspace{-15pt}
\end{figure*}
Within switching-mode power supplies, capacitors and semiconductor power switches constitute the most common sources of failure, whereas faults associated with resistors and inductors occur less frequently \cite{kumar2020review}. Capacitor failures typically originate from material wear-out mechanisms, operation beyond specified limits, or manufacturing defects. Power-switch failures are predominantly induced by excessive thermal transients or electrical overstress events and commonly manifest as either open-circuit faults (OCFs) or short-circuit faults (SCFs), with SCFs generally posing more severe system-level risks \cite{kumar2020review}. The impact of a component-level fault on overall system performance depends on both the fault type and the physical location of the affected device within the DVPD network.

A primary focus of this paper is on semiconductor device failures that manifest as short-circuit and open-circuit faults caused by elevated electrical and thermal stresses associated with dense integration and vertical stacking of VR components within the substrate. A fault-detection and isolation methodology is proposed for high-performance DVPD networks. To evaluate the system-level impact of power-switch faults, a representative DVPD architecture is analyzed using a combination of circuit-level and system-level simulations. The VR power stages, shared PDN, and current loads are designed and simulated in SPICE to accurately capture switching behavior, parasitic effects, and fault dynamics under both healthy and faulted operating conditions. The fault-detection logic, inductor-current emulator model, isolation control, and signal-processing blocks are implemented in MATLAB/Simulink. Parameters and operating conditions are cross-verified between the SPICE and MATLAB/Simulink simulations to ensure consistency. This simulation framework enables quantitative assessment of fault severity and detection latency, which directly informs the proposed fault-detection and isolation strategy.

The effect of power-switch short-circuit and open-circuit faults on system-level efficiency is examined in Section~\ref{sec:efficiency}. The proposed real-time fault-detection and isolation methodology is presented in Section~\ref{sec: detection}, and conclusions are provided in Section~\ref{sec: conclusion}.

\section{Impact of Faults on System Performance}\label{sec:efficiency}

Consider a DVPD system comprising multiple distributed synchronous buck or buck-derived VRs (e.g., \cite{kirshenboim2017high-efficiency}) that collectively supply spatially distributed current loads, as illustrated in Fig.~\ref{fig:PDN}. A high-side SCF (HS-SCF) in a VR results in continuous inductor charging, which elevates the local output node and, consequently, the shared low-voltage power plane (typically 1~V or 0.9~V) toward the input rail (e.g., 6~V, 12~V or 48~V). Such an overvoltage condition is prohibitive for downstream loads, which are typically designed to operate at sub-1~V levels. Conversely, a low-side SCF (LS-SCF) forces continuous inductor discharge by clamping the switching node to ground. This behavior diverts current from neighboring VRs, disrupts current sharing, and degrades voltage regulation across the shared power plane. In both cases, SCFs establish low-impedance paths between the supply and ground, resulting in excessive short-circuit power dissipation and threatening system-wide stability and reliability.

In contrast, OCFs exhibit less severe system-level consequences. A high-side OCF (HS-OCF) disconnects the inductor from the input supply, causing a gradual decay of the local output voltage as the inductor discharges into the load. This voltage droop increases the current demand on the remaining VRs, thereby elevating their electrical and thermal stress. A low-side OCF (LS-OCF) disables the synchronous conduction path to ground, forcing the inductor current to freewheel through the body diode of the corresponding power switch. This condition introduces additional conduction loss and a moderate output voltage droop, which is typically compensated by the closed-loop controller through duty-cycle adjustment. As a result, OCFs impose a comparatively modest impact on VR efficiency and overall system performance. 

\begin{figure}[b!]
  \vspace{-20pt}
  \centerline{\includegraphics[width=0.4\textwidth]{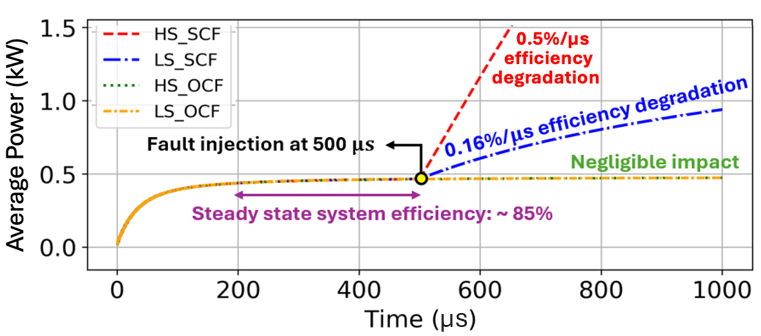}}
  \vspace{-10pt}
  \caption{Impact of SCF and OCF in power switches on system-wide efficiency}
  \label{fig:FaultImpact}
\end{figure}

To illustrate the impact of each fault type on DVPD, consider a representative system, comprising 20 buck regulators stepping down 6~V to 1~V, delivering up to 400~W, and operating at a switching frequency of 1~MHz. The system is designed and simulated in SPICE. The average power consumption of the system under the considered fault scenarios is shown in Fig.~\ref{fig:FaultImpact}.
The allowable fault-detection latency is governed by the rate of system-wide efficiency degradation following a fault event. As illustrated, HS-SCFs and LS-SCFs exhibit efficiency degradation rates of approximately 0.5$\frac{\%}{\text{\textmu} s}$ and 0.16$\frac{\%}{\text{\textmu} s}$, respectively. These degradation rates dictate detection and isolation time requirements of 10~\textmu s for HS-SCFs and 31~\textmu s for LS-SCFs to maintain system efficiency above 80\% under all operating conditions.

\section{Faults Detection and Isolation}\label{sec: detection}
Existing fault-detection methodologies are primarily developed for industrial and automotive systems that employ discrete VRs with relaxed area constraints, delivering up to 100~W at 3.3~V operating at switching frequencies below 300~kHz~\cite{poon2016model,ashourloo2020fault,pazouki2017fault}. These approaches typically rely on additional diagnostic hardware, which introduces extra cost and area overhead. Such methods are insufficient for VRs integrated under stringent area constraints and operating in the megahertz range while collectively delivering load power approaching 1~kW at 1~V.

To address this challenge, a fault-diagnosis module that employs voltage and current sensors and an inductor-current emulator is proposed. The emulator analytically predicts the inductor current at a given time based on the inductor current from the previous switching cycle, the input voltage, the output voltage obtained from sensor measurements, and the duty cycle provided by the controller. Note that all measurements required by the fault-diagnosis module are already available from sensors within the converter as part of the control circuitry, and no additional hardware is required.

\subsection{Current Emulator}\label{subsec:CurrentEmulator}
An analytical model is developed to estimate the inductor current in a voltage regulator based on standard inductor current behavior \cite{chen2003predictive}. Consider a buck or buck-derived voltage regulator operating in continuous conduction mode (CCM). When the high-side (HS) switch is turned on, one terminal of the inductor is connected to the input supply while the other is connected to the converter output. During this interval, the inductor current increases with a slope given by 
\begin{equation}
  \Delta i_L^{HS} = \frac{(V_{in}-V_o)D}{L f_{sw}},
  \label{eq:IL_HS}
\end{equation}
where $V_{in}$ and $V_o$ are the converter input and output voltages, $D$ is the duty cycle, $L$ is the inductance, and $f_{sw}$ is the converter switching frequency.
Conversely, when the LS switch is turned on, the inductor discharges into the load, causing the current to decrease with a slope
%
\begin{equation}
  \Delta i_L^{LS} = \frac{-V_o(1-D)}{L f_{sw}}.
  \label{eq:IL_LS}
\end{equation}

Assuming that the inductor current is sampled at a rate of ${K \times f_{sw}}$, the instantaneous inductor current at the $n^\text{th}$ sampling instant, $I_L^{\mathrm{est}}(n)$, is estimated based on 1) the inductor current measured during the previous switching cycle (i.e., $K$ samples earlier), $I_L(n-K)$, 2) expected first-order resistive damping of the inductor current over one switching period, and 3) the analytical rate-of-change of the inductor current, $\Delta i_L^{HS}-\Delta i_L^{LS}$, as given by (\ref{eq:IL_HS}) and (\ref{eq:IL_LS}). The estimated inductor current at sample $n$ is therefore
\begin{equation}
\small
    I_L^{\mathrm{est}}(n) = I_L(n-K)\left(1-\frac{R^\text{eff}}{Lf_{sw}}\right)
    + \frac{V_{in}(n) D(n) - V_o(n)}{L f_{sw}},
  \label{eq:IL_est}
\end{equation}
where $R^\text{eff} = R_L + R_{HS}D(n) + R_{LS}(1-D(n))$ is the effective series resistance seen by the inductor over the switching cycle, comprising the inductor series resistance ($R_L$) and averaged HS ($R_{HS}$) and LS ($R_{LS}$) conduction resistances.

\begin{figure}[t!]
  \centerline{\includegraphics[width=0.4\textwidth]{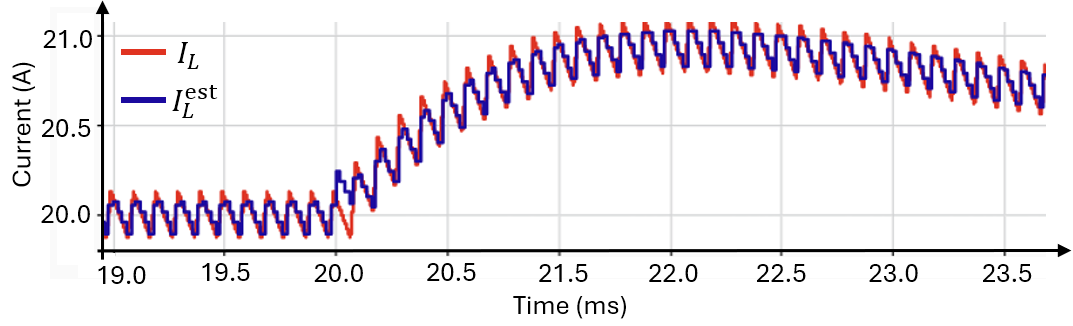}}
  \vspace{-10pt}
  \caption{Measured and estimated inductor current under transient conditions induced by a step change in input voltage from 4~V to 5~V at 20~ms, while delivering 20~A of load current at a regulated output voltage of 1~V.}
\label{fig:CurrentEmulatorResult}
\vspace{-15pt}
\end{figure}

To evaluate the proposed current model, a representative buck converter is designed and simulated in SPICE. The inductor response is analyzed under transient conditions induced by a step change in the input voltage from 4 V to 5 V at 20 ms, while delivering a 20 A load current at a regulated 1 V output and 1 MHz.
A comparison between the estimated and simulated inductor current is shown in Fig.~\ref{fig:CurrentEmulatorResult}, exhibiting less than 0.2\% error in steady state. Note that the model realigns with the accurate current value within a single switching cycle following the input voltage step.

\subsection{Fault Detection}\label{subsec:FaultDetection}

\begin{figure*}[t!]
\centerline{\includegraphics[width=0.97\textwidth]{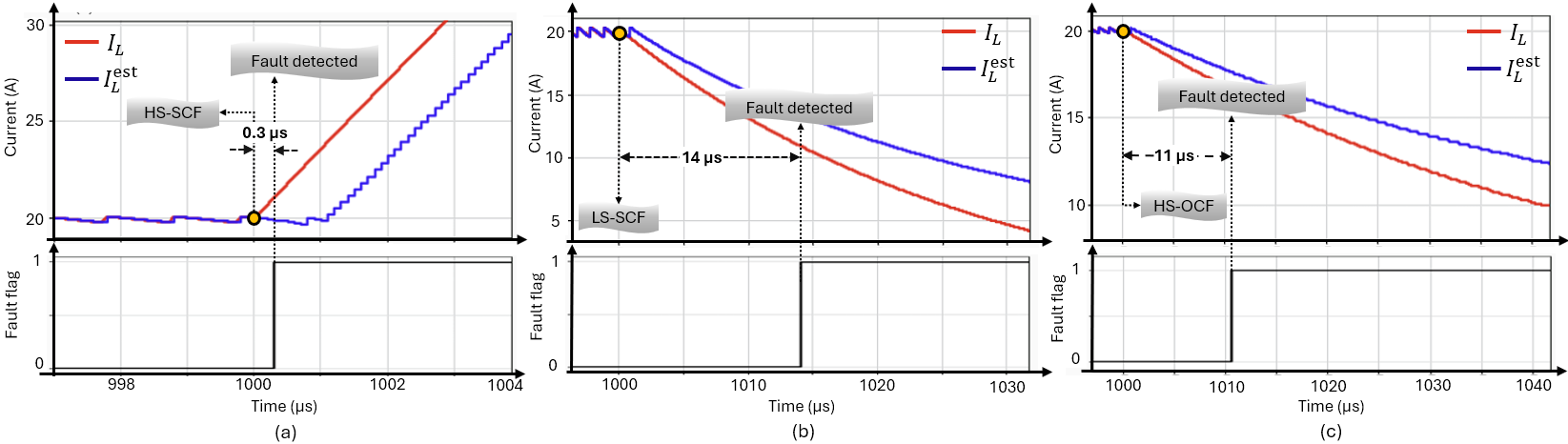}}
\vspace{-10pt}
\caption{Illustration of the impact of different fault types on the inductor current and the corresponding fault-flag signal generated by the proposed detection framework for (a) HS-SCF, (b) LS-SCF, and (c) HS-OCF.}
\label{fig:FaultDetection}
\vspace{-15pt}
\end{figure*} 

Upon the occurrence of a fault, the inductor current deviates from normal operating behavior. Depending on the fault type and the affected switch, the inductor current may exhibit a monotonic increase, or a monotonic decrease, rather than the steady-state DC behavior with finite ripple observed during normal operation. This deviation provides a distinct electrical signature for fault detection, which is identified by comparing the measured inductor current, $I_L(n)$, with the estimated inductor current, $I_L^{\mathrm{est}}(n)$, at each sampling instant. A fault is flagged when the absolute error between the measured and estimated inductor currents exceeds a threshold, $\tau(n)$, 
\begin{equation}
  \epsilon(n) = \left| I_L^{\mathrm{est}}(n) - I_L(n) \right| > \tau(n).
  \label{eq:Error}
\end{equation}
The detection threshold at each sampling instant, $\tau(n)$, is defined in this paper as 5\% of the inductor current measured during the previous switching cycle, ${I_L(n-K)}$.

To demonstrate the proposed fault-detection methodology, a buck converter is designed to step down a 6~V input voltage to a regulated 1~V output, switching at 1~MHz, while delivering up to 20~A of load current. Faults are injected at 1~ms to evaluate the detection performance under controlled conditions. Fault detection for HS-SCFs, LS-SCFs, and HS-OCFs is illustrated in Fig.~\ref{fig:FaultDetection}. 
Note that an increase or decrease in inductor current alone is not sufficient to indicate a fault. Under normal operating conditions, changes in inductor current are expected and are consistent with variations in the input voltage, load current, or duty cycle. Under faulty conditions, however, the observed current behavior violates these expected relationships, as the rate of change of the inductor current is no longer consistent with the known circuit variables.
The proposed method detects HS-SCFs within 0.3~\textmu s, LS-SCFs within 14~\textmu s, and HS-OCFs within 11~\textmu s. These detection latencies satisfy the detection and isolation speed requirements derived in Section~\ref{sec:efficiency}. The  proposed fault-detection scheme therefore enables fast and reliable identification of both short-circuit and open-circuit fault scenarios, while maintaining sufficient time margin to isolate faulty VRs before fault-induced degradation propagates through the shared PDN.

\subsection{Fault Isolation}

\begin{figure}[b!]
\vspace{-15pt}
\centerline{\includegraphics[width=0.4\textwidth]
{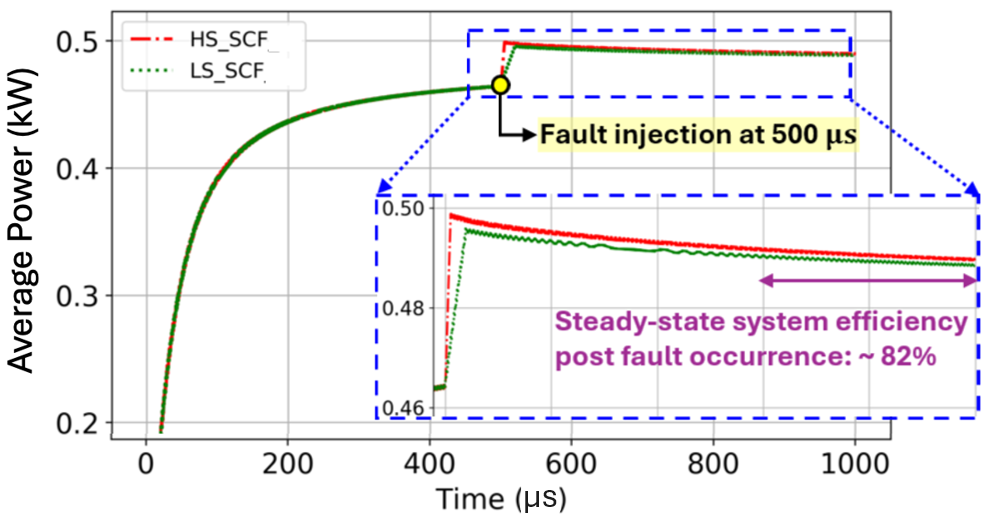}}
\vspace{-10pt}
\caption{Impact of SCFs on fault-torrent DVPD system performance }
\label{fig:FaultIsolation}
\end{figure}

To mitigate the impact of faults and prevent fault propagation, each VR in the DVPD system is equipped with a fault-detection module incorporating an inductor-current emulator and two fuse-based isolation elements. One fuse is placed in series with HS switch at the converter input, while a second fuse is placed at the converter output (see Fig. \ref{fig:PDN}). Upon fault occurrence, the asserted fault-flag signal triggers both protection fuses. The input-side fuse disconnects the supply from the faulty VR, thereby eliminating the supply-to-ground low-impedance current path, while the output-side fuse electrically decouples the faulty VR from the shared PDN. This dual-fuse isolation strategy provides complete electrical isolation of the faulted regulator from both the supply and the power delivery network, allowing the remaining healthy VRs to continue supplying the load without interruption.

The proposed methodology is evaluated on the same DVPD system described in Section~\ref{sec:efficiency}, which serves to quantify the impact of faults on system-wide efficiency, as illustrated in Fig.~\ref{fig:FaultImpact}. Performance of the system designed with the proposed fault-detection and isolation methodology is illustrated in Fig.~\ref{fig:FaultIsolation}. Based on the results, the adverse effects of SCFs are effectively mitigated through the proposed fault-detection and isolation scheme. Following isolation of the faulty VR, the remaining 19 VRs continue to supply the 400~W load, resulting in a modest reduction in system-wide efficiency from 85\% to 82\% due to increased current stress on the healthy regulators. In contrast, OCFs inherently result in self-isolation of the faulty VR from the PDN, thereby redistributing the load current among the remaining healthy VRs.


\section{Conclusion}\label{sec: conclusion}
A real-time fault-detection and isolation methodology for DVPD architectures is presented in this paper, including the derivation of latency requirements for fault detection and isolation and the effective identification and isolation of faulty VRs. It is important to note that operation at megahertz switching frequencies, as enabled by state-of-the-art VRs, requires signal sampling rates on the order of several megahertz. In practice, real-time signal-processing constraints in digital controllers may introduce additional latency and bandwidth limitations that affect fault detection and isolation. Investigation of these implementation considerations, along with experimental validation of the proposed framework, is identified as an important direction for future work.

\bibliographystyle{IEEEtran}
\bibliography{./master_bib}

\end{document}